\documentclass[12pt]{article}
\usepackage{epsf}
\usepackage{graphicx}
\usepackage{eepic}
\usepackage[colorlinks]{hyperref}
\usepackage{cite}
\usepackage{color}
\usepackage{latexsym,epsfig}
\newcommand{\be}{\begin{equation}}
\newcommand{\ee}{\end{equation}}
\newcommand{\ba}{\begin{array}{c}}
\newcommand{\ea}{\end{array}}
\newcommand{\bqa}{\begin{eqnarray}}
\newcommand{\eqa}{\end{eqnarray}}

\newcommand{\no}{\nonumber\\}
\begin{document}

\begin{center}
{\Large\bf Ambiversion of  X(3872) }
\\[10mm]
{\sc Ou Zhang,$^1$ C.~Meng,$^1$ H.~Q.~Zheng$^2$ }
\\[2mm]
\end{center}
{\it 1: Department of Physics, Peking University, Beijing 100871,
China}\\
 {\it  2:  Department of Physics and State Key Laboratory of
Nuclear Physics and Technology, Peking University, Beijing 100871,
China}
\\[5mm]
\begin{center}
\today

\end{center}

\vskip 1cm
\begin{abstract}
An analysis including most recent Belle data on $X(3872)$ is
performed, using coupled channel Flatt\'e formula. A third sheet
pole close  to but \textit{ below} $D^0D^{*0}$ threshold is found,
besides the bound state/virtual state pole discussed in previous
literature. The co-existence of two poles near the $D^0D^{*0}$
threshold indicates that the $X(3872)$  may be of ordinary $c\bar c$
$2 ^3P_1$ state origin, distorted by strong coupled channel effects.
The latter manifests itself as a molecular bound state (or a virtual
state).
\end{abstract}


\section{Introduction}

In year 2003 the Belle collaboration found a very narrow
($\Gamma_X<2.3$ MeV) resonance structure named X(3872) in the
$J/\Psi\,\pi\pi$ invariant mass spectrum, in the $B^+\to
K^+J/\Psi\,\pi^+\pi^-$ process~\cite{Belle1}. The branching ratio $
\mathrm{Br}(B^+\to K^+ X)\,\mathrm{Br}(X\to \pi^+\pi^- J/\Psi)$ is
updated to be $=(7\mbox{-}10)\times 10^{-6}$ both by
BaBar~\cite{BaBar08:X3872:R_(0/+)} and by
Belle~\cite{Belle08:X3872:R_(0/+)}. Moreover, Belle also observed
$X(3872)$ in the $B^0$ decay and found that the rate is comparable
with that of the charged channel~\cite{Belle08:X3872:R_(0/+)}. Most
recently, the CDF Collaboration reported a new measurement on the
mass parameter in the $J/\Psi\pi^+\pi^-$
channel~\cite{CDF08:X3872:MASS},
 \be
M_X= 3871.61\pm 0.16\pm 0.19\,\mbox{MeV}\label{M:X3872-CDF}\ . \ee
Replacing the old CDF measurement by the new one results in a world
average of $M_X= 3871.51 \pm 0.22$MeV, which is very close to the
$D^0\bar{D}^{*0}$ threshold
$M_{D^0\bar{D}^{*0}}=3871.81\pm0.36$~\cite{Cleo07:D-mass}.

The other decay modes of $X(3872)$ include
$J/\Psi\pi^+\pi^-\pi^0$~\cite{Belle05:X3872:psi-omega},
$J/\Psi\gamma$~\cite{BaBar08:X3872:psi(')-gamma} and
$\Psi'\gamma$~\cite{BaBar08:X3872:psi(')-gamma} with relative rates
\bqa\label{pipipiJPsi}R\equiv
\frac{\mathrm{Br}(X\to \pi^+\pi^-\pi^0 J/\Psi)}{\mathrm{Br}(X\to
\pi^+\pi^- J/\Psi)}&=&1.0\pm 0.5\ ,\label{22}\\
\frac{\mathrm{Br}(X\to \gamma
J/\Psi)}{\mathrm{Br}(X\to \pi^+\pi^- J/\Psi)}&=&0.33\pm 0.12,\\
\frac{\mathrm{Br}(X\to \gamma \Psi')}{\mathrm{Br}(X\to \pi^+\pi^-
J/\Psi)}&=&1.1\pm 0.4\,. \label{psiprime-gamma} \eqa
The dipion mass spectrum in the $J/\Psi\pi^+\pi^-$ mode shows that
they come from the $\rho$ resonance~\cite{Belle1} and the 3$\pi$ in
the $J/\Psi\pi^+\pi^-\pi^0$ mode come from the $\omega$
resonance~\cite{Belle05:X3872:psi-omega}. Thus, the rario $R\simeq1$
in (\ref{22}) indicates that there should be large isospin violation
in the decays of $X(3872)$.

In year 2006 the Belle collaboration studied the $B^+\to
D^0\bar{D}^0\pi^0 K^+$ decay process and found the enhancement of
the $D^0\bar D^0\pi^0$ signal just above the $D^0\bar {D}^{*0}$
threshold~\cite{Belle06:X3872:DDpi}, the resonance is peaked at
 \be\label{Bellepole2} M_X=3875.2
\pm 0.7^{+0.3}_{-1.6}\pm 0.8\,\mbox{MeV},\ee roughly 3.6 MeV higher
than the value in (\ref{M:X3872-CDF}). The corresponding branching
ratio at the $D^0\bar{D}^0\pi^0$ peak is~\cite{Belle06:X3872:DDpi},
\be \mathrm{Br}(B^+\to K^+D^0\bar{D}^0\pi^0)=(1.02\pm
.31^{+0.21}_{-0.29})\times 10^{-4}\ .\ee The different peak
locations of X(3872) in the $D^0\bar D^0\pi^0$ ($D^0\bar D^{*0}$)
and $J/\Psi\pi^+\pi^-$ channels are reconfirmed by latter BaBar
experiments~\cite{BaBar08:X3872:DD*}. In 2008, a new analysis to the
Belle data in the $D^{*0}\bar D^0$ ($D^{*0}\to D^0\pi^0$ and
$D^{*0}\to D^0\gamma$) channel is given~\cite{Belle08:X3872:DD*},
and the new determination of the peak  is 2.6MeV lower than the
previously reported by Belle~\cite{Belle08:X3872:DD*},
 \be\label{valueDDpi2}
 M_X= 3872.6^{+0.5}_{-0.4}\pm 0.4\mbox{MeV}\ .
 \ee
The difference comes from the inclusion of new data ($D^*\to
D\gamma$), more sophisticated fit (unbinned fit with mass dependent
resolution), and improved Breit--Wigner formula (the Flatt\'e
formula). The central value as given by Eq.~(\ref{valueDDpi2}) is,
however, still about 1MeV above than the value measured by CDF
Collaboration~\cite{CDF08:X3872:MASS}. Meanwhile in
Ref.~\cite{Belle08:X3872:DD*} a renewed determination of the
following branching ratio is given, \be \mbox{Br}(B^+\to
K^+X(D^{*0}\bar D^0))=(0.73\pm 0.17\pm 0.13)\times 10^{-4}\ .
 \ee

The $X(3872)$ is naturally interpreted as a $C=+$ molecule of
$D^0\bar{D}^{*0}$ in
$s$-wave~\cite{Tornqvist:X3872:molecule,Swanson04:X3872:molecule}
since its mass is very close to the $D^0\bar{D}^{*0}$ threshold and
the quantum number $J^{PC}=1^{++}$ is favored by the experimental
analysis~\cite{Belle05CDF06:X3872:J^PC}. It also predicted the
$J/\Psi\omega$ mode with similar rate as
$J/\Psi\rho$~\cite{Swanson04:X3872:molecule}. However, the large
production rates of $X(3872)$ in B-factories and at Tevatron favor a
conventional charmonium
assignment~\cite{Meng05:X3872:B-production,suzuki05:X3872:chi-c1-2P}
(say, $\chi_{c1}'$) rather than a loosely bound state of
$D^0\bar{D}^{*0}$. Furthermore, the large decay rate of
$X\to\Psi'\gamma$ in (\ref{psiprime-gamma}) also strongly disfavors
the molecular assignment since it is very difficult for the
transition of a molecular to $\Psi'$ through the quark annihilation
mechanism~\cite{Swanson04:X3872:molecule}. The large isospin
violation indicated by (\ref{22}) can also be explained quite well
in the charmonium model~\cite{Meng07:X3872:decays}. Hence it seems
that we are facing a dilemma in recognizing $X(3872)$.

To further clarify the identity of $X(3872)$, one needs to look
deeper into the pole structures of the scattering amplitude
involving $X(3872)$. For a dynamical molecule of $D^0\bar{D}^{*0}$,
there is only one pole near the threshold, and the requirement of
two nearby poles to describe the $X(3872)$ will generally imply that
it is a $c\bar{c}$ state near the
threshold~\cite{Morgan92:pole-counting}. The line shapes of $B^+\to
XK^+$ in the $J/\Psi\pi^+\pi^-$ and
$D^0\bar{D}^0\pi^0/D^0\bar{D}^{*0}$ modes and the corresponding pole
structures have been studied by two groups~\cite{hanhart,braaten}
independently. Both fits give an one-pole structure, although one
fit~\cite{hanhart} favors a virtual state and the
other~\cite{braaten} favors the loosely bound state. Since more data
are available after these two fits, it deserves a careful reanalysis
 to the data of $X(3872)$. In
this paper we  devote to the study of this problem. In Sec.~\ref{2},
we firstly describe the method we use for the analysis, we also
describe how we make the fit from various experimental data. Sec. 3
is for the discussions and conclusions. The final result of this
analysis presents a unified picture in understanding the dual faces
of $X(3872)$: A third sheet pole close  to but \textit{ below}
$D^0D^{*0}$ threshold is found, besides the bound state/virtual
state pole discussed in previous literature. The co-existence of two
poles near the $D^0D^{*0}$ threshold indicates that the $X(3872)$
may be of ordinary $c\bar c$ $2 ^3P_1$ state origin, distorted by
strong coupled channel effects. The latter manifests itself as a
molecular bound state (or a virtual state).

\section{Coupled channel description of the X(3872)
resonance}\label{2}



Notice that $X(3872)$ associates with nearby different cuts, hence a
coupled channel analysis is needed in order to take care of the
complicated singularity structure. This has already been emphasized
in Refs.~\cite{hanhart,braaten,bugg,oset}. Hanhart et al. gave a
very interesting explanation to the X(3872) peak as a virtual
state~\cite{hanhart}. Their conclusion relied on of course the
experimental data available, and especially on the two peak
structure in different channels. The latter plays a crucial role in
getting such a conclusion. In the analysis of Hanhart et al., the
effect of energy resolution is not considered. Since the two peaks
are not too far from each other and the difference between them is
comparable in magnitude to the energy resolution parameter, one
worries about that the negligence of energy resolution effect may
distort their conclusion. For reasons mentioned previously a new
analysis on this subject is necessary. We proceed with data prsently
available~\cite{BaBar08:X3872:R_(0/+),Belle08:X3872:R_(0/+),BaBar08:X3872:DD*,Belle08:X3872:DD*}
to reanalyze the $X(3872)$ state, with the energy resolution effect
taken into account. On the theory side the method we use is
essentially the same as that of Ref.~\cite{hanhart}.


For describing the chain decays with X(3872) involved as
intermediate state, we parameterize the inverse of the propagator of
X(3872) as \be\label{XBreit_wigner}
D(E)=E-E_f+\frac{i}{2}(g_1k_1+g_2k_2+\Gamma(E)+\Gamma_c)\ , \ee
where {$E_f=M_X-M_{D^0}-M_{\bar{D}^{*0}}$}; $k_1=\sqrt{2\mu_1E}$,
$k_2=\sqrt{2\mu_2(E-\delta)}$ and $\delta=M_{D^+}+M_{D^{*-}}
-M_{D^0}-M_{\bar{D}^{*0}}$, $\mu_1$ and $\mu_2$ are the reduced
masses of  $D^0\bar{D}^{*0}$ and $D^+{D}^{*-}$, respectively.
Isospin symmetry requires $g_1\simeq g_2$. $\Gamma(E)$ includes
channels  $J/\Psi\pi^+\pi^-$ (through $J/\Psi\,\rho$),
$J/\Psi\pi^+\pi^-\pi^0$ (through $J/\Psi\,\omega$) channels.
Different from Ref.~\cite{hanhart} here we add a constant width
$\Gamma_c$ to simulate every other  channels, including radiative
decays and light hadron decays. From Eq.~(\ref{psiprime-gamma}) we
know that this term is certainly non-negligible as comparing with
$J/\Psi\pi^+\pi^-$ decay, not to mention the to be observed light
hadronic decays.

For simplicity, we describe the $\rho$ and $\omega$ resonances in
the final states by their Breit-Wigner distribution functions, then
one has \bqa
\Gamma(E)&=&\Gamma_{\pi^+\pi^-J/\Psi}(E)+\Gamma_{\pi^+\pi^-\pi^0
J/\Psi}(E)\ ,\no \Gamma_{\pi^+\pi^-J/\Psi}(E)&=&
f_\rho\int^{M_X-m_{J/\Psi}}_{2m_\pi}\frac{dm}{2\pi}\frac{k(m)\Gamma_\rho}
{(m-m_\rho)^2+\Gamma^2_\rho/4}\ ,\no \Gamma_{\pi^+\pi^-\pi^0
J/\Psi}(E)&=&f_\omega\int^{M_X-m_{J/\Psi}}_{3m_\pi}\frac{dm}{2\pi}\frac{k(m)\Gamma_\omega}
{(m-m_\omega)^2+\Gamma^2_\omega/4}\ , \eqa where $f_\rho$ and
$f_\omega$ are the $X$ couplings to $J/\Psi\rho$ and $J/\Psi\omega$
respectively, $M_X=E+M_{D^0}+M_{\bar{D}^{*0}}$ is the (off-shell)
center of mass energy of the X particle and \be
k(m)=\sqrt{\frac{(M^2_X-(m+m_{J/\Psi})^2)(M^2_X-(m-m_{J/\Psi})^2)}{4M^2_X}}\
. \ee

Let ${\cal B}=Br(B\to XK)$, recalling that $Br(D^{*0}\to
D^0\pi^0)=61.9\pm 2.9\%$,~\cite{PDG08} repeatedly using the chain
decay formulae leads to,
 \bqa\label{bran_DDpi_Jpsipipi_Jpsiomega}
\frac{d\mathrm{Br}[B\to K D^0\bar D^{0}\pi^0]}{dE}&=&0.62{\cal
B}\frac{1}{2\pi}\frac{\Gamma_{D^0\bar D^{*0}}(E)} {|D(E)|^2}\ ,\no
\frac{d\mathrm{Br}[B\to K\pi^+\pi^-J/\Psi]}{dE}&=&{\cal
B}\frac{1}{2\pi}\frac{\Gamma_{\pi^+\pi^-J/\Psi}(E)} {|D(E)|^2}\ ,\no
\frac{d\mathrm{Br}[B\to K \pi^+\pi^-\pi^0 J/\Psi]}{dE}&=&{\cal
B}\frac{1}{2\pi} \frac{\Gamma_{\pi^+\pi^-\pi^0 J/\Psi}(E)}
{|D(E)|^2}\ . \eqa In the fit to $X\to\bar D^{*0}D^0$
data~\cite{BaBar08:X3872:DD*}, since all decay modes of $D^{*0}$ are
considered there, we drop the factor 0.62 in the first formula of
the above equation.

One also has to consider the background contributions. In all the
fit to the data, we assume there is no interference between data and
background. This is in coincidence with experimental analyses. In
$DD\pi$ channel we assume the  background contribution is
proportional to $E_{DD\pi}$, hence
 \be\label{bgcase1}
\frac{d\overline{\mathrm{Br}}[B\to K D^0\bar
D^{0}\pi^0]}{dE}=0.62{\cal B}\frac{1}{2\pi}\frac{gk_1} {|D(E)|^2}+
c_{b.g.}E_{DD\pi}\ . \ee For the $D^{*0}D^0$ final state  we assume
background contribution is proportional to the phase space of
$D^0\bar{D}^{*0}$, $k_1$. In the $J/\Psi\pi^+\pi^-$ case, we assume
the background is a constant. Herewith  we often use overlined
branching ratios to represent the signal plus background
contributions:
 \be
\frac{d\overline{\mathrm{Br}}(E)}{dE}=\frac{d\mathrm{{Br}}(E)}{dE}+
{b.g.}(E)\ . \ee

The ratio $R$ defined in Eq.~(\ref{pipipiJPsi}) has to be put into
the fitting program as a constraint. Throughout this paper the ratio
$R={Br(X\to J/\Psi\rho)\over Br(X\to J/\Psi\omega)}$ is set to 1.
The formula used to estimate $R$  is the same as that adopted by
Hanhart et al.~\cite{hanhart}, but in here we constrain the value
$R$ by using the penalty function method, which is simple and
effective. That is we effectively add a term to the total  $\chi^2$:
$\chi^2_{R}=10\times {|R-1|^2}/{0.4^2}$. The factor 10 is an
arbitrarily chosen  penalty factor which is enough to make the ratio
$R$ being almost exactly unity. We notice that the ratio $R$
measured by experiments contains a large error bar as shown in
Eq.~(\ref{22}).   We will therefore also pay some attention in the
numerical fit to different value of $R$, in next section.

\section{The data fitting program and the fit results}
\subsection{Data samples  and the energy resolution parameters}
 As stated earlier we use 4 sets of data:
\begin{enumerate}
\item [1:] The $X\to\bar D^{*0}D^0$ mode by BaBar~\cite{BaBar08:X3872:DD*}, where
$\bar D^{*0}$ is reconstructed both from $D^0\pi^0$ and $D^0\gamma$
mode. There are 12 data points in the fit region from $\bar
D^{*0}D^0$ threshold up to 3.895GeV. The background contribution
starts from $\bar D^{*0}D^0$ threshold, the same as that adopted in
Ref.~\cite{BaBar08:X3872:DD*}. The corresponding  number of events
distribution is,
 \be\label{Norm1}
 N^{D^0D^{*0}}_{BaBar}=2\mbox{[MeV]}\,\frac{33.1}{1.67\times 10^{-4}}\,
  \frac{d\overline{\mathrm{Br}}[B\to K D^0D^{*0}]}{dE} .
 \ee

 \item [2:]  The $X\to D^0\bar D^0\pi^0$ data from Belle
 Collaboration~\cite{Belle06:X3872:DDpi} is replaced by the upgraded one
 from $B^\pm\to XK^\pm$~\cite{Belle08:X3872:DD*}.
We fit the data   in the energy region from   $D^0D^0\pi^0$
threshold to 3.91257GeV, there are totally 119 events collected from
$B^{\pm}$ decays.

\item [3:] Data of $J/\Psi\pi^+\pi^-$   from
BaBar~\cite{BaBar08:X3872:R_(0/+)}.  We only use the charge mode
($B^+\to X(3872)K^+$) data, since the error bar of the neutral mode
($B^0\to X(3872)K^0$) data are much larger. There are 11 data points
in the fit region $3.84<M_X<3.89$GeV and
 \be\label{Norm4}
 N^{J/\Psi\pi^+\pi^-}_{BaBar}=5\mbox{[MeV]}\,\frac{93.4}{8.4\times 10^{-6}}\,
 \frac{d\overline{\mathrm{Br}}[B\to K J/\Psi\pi^+\pi^-]}{dE} .
 \ee

 \item [4:] Data of $J/\Psi\pi^+\pi^-$   from
most recent Belle experiments~\cite{Belle08:X3872:R_(0/+)}.
 We fit
the data sample in the energy region from 3.84135GeV to 3.90173GeV,
with  totally {398} events.
\end{enumerate}

 With the unbinned data sets from Belle
Collaboration on both $X\to D^0D^0\pi^0$ from $B^\pm\to XK^\pm$
  and  $X\to J/\Psi\pi^+\pi^-$ decay, we make a combined fit of
  likelihood method and $\chi^2$ method in
the following way:
 \be\label{chisquareeffective}
  \chi^2_{eff}\equiv -2\sum_i\log{\cal L}_i+\sum_j\chi^2_j+\chi_R^2\ ,
\ee where $i=\mbox{2}, \mbox{4}$; $j=\mbox{1}, \mbox{3}$.  The
background contributions to  the two data samples 2 and 4 are
treated similarly as those discussed previously. The PDF used in the
likelihood fit is written as,
 \be \mu
(E)=\frac{\frac{d\,\mathrm{Br}(E)}{dE}+b.g.}{\int dE \left[{d
{\mathrm{Br}}(E)\over dE}+b.g.\right]}\ .
 \ee

Because the peaks in different channels are rather close to each
other, one needs to take energy resolution effect into account,
 \be
Br(E)=\int dE_x \,
Br(E_X)\,{e^{-\frac{(E_x-E)^2}{2\sigma(E_x)^2}}\over
\sqrt{2\pi}\sigma(E_x)}\,  \ . \ee
 In general, the
energy resolution parameter $\sigma$ is a function of $E_x$, the
original energy of incoming particles. For $J/\Psi\pi^+\pi^-$
channel at Belle: $ \sigma(E_x)=3 MeV\ .$ For $D^{*0}D^0$ at Belle:
$ \sigma(E_x)\simeq 0.176\sqrt{E_x-M_{D^{*0}D^0}}\
.$~\cite{Belle08:X3872:DD*} We assume that the BaBar detector
maintains the same energy resolution parameters.

\subsection{Pole locations determined from combined data fit}
\label{unbinned}

Experiments indicate ${\cal B}$ to be about a few times $10^{-4}$.
The value of ${\cal B}$ is about $2\mbox{-}4\times10^{-4}$ in the
charmonium model~\cite{Meng05:X3872:B-production}, while in the
molecular model, it is in general not larger than
$1\times10^{-4}$~\cite{braaten,Braaten:X3872:B-production}.
Therefore in the following analyses, we often fix $\cal B$ at a few
times $10^{-4}$, though it is noticed that the fit program prefers a
larger value of ${\cal B}\sim 2\times 10^{-3}$ with large error
bars.

We have stressed in section~\ref{2} that we add a constant width
term $\Gamma_c$ in the Flatt\'e propagator, which corresponds to
modes rather than the near threshold ones ($J/\Psi\,\rho,\
J/\Psi\,\omega,\ DD\pi$). These modes include both the observed
ones, such as $\Psi^{(\prime)}\gamma$
\cite{BaBar08:X3872:psi(')-gamma}, and the hidden ones. In the
charmonium model, the most important hidden decay mode of $X(3872)$
as $\chi_{c1}(2P)$ is the inclusive light hadronic decay, and the
partial width is of $\mathcal {O}(1)$
MeV~\cite{Meng07:X3872:decays}. However, for the pure $D^0\bar
D^{*0}$ molecule, it is difficult to annihilate the charm quark pair
into light hadrons. Therefore, the most important hidden modes of
$X(3872)$ in the molecular model may be the hadronic transitions to
$\chi_{cJ}(1P)$, such as $\chi_{c0}\pi^0$ and $\chi_{c1}\pi\pi$,
while the widths of them are expected to be smaller than that of
$J/\Psi\pi^+\pi^-$~\cite{Molecule:XtoChicJ}. Thus, the term
$\Gamma_c$ can provide important information on the $X(3872)$.

Poles on different sheets are searched for using results of fit
parameters. The naming scheme of Riemann sheets is given in
table~\ref{tab00}.
\begin{table}[htp]
\begin{center}
\doublerulesep 0pt
\renewcommand\arraystretch{1.1}
\begin{tabular}{|c|c|c|c|}
\hline

  & II  &  III   &  IV   \\
\hline $\Gamma(E)+\Gamma_c$ & $\,\,\,\,\,\,-\,\,\,\,\,\,$  &
$\,\,\,\,\,\,-\,\,\,\,\,\,$ & $\,\,\,\,\,\,+\,\,\,\,\,\,$
    \\
\hline
 $g_1k_1$ & $\,\,\,\,\,\,+\,\,\,\,\,\,$  &  $\,\,\,\,\,\,-\,\,\,\,\,\,$  &  $\,\,\,\,\,\,-\,\,\,\,\,\,$     \\
\hline
\end {tabular}
\caption {\label{tab00}Naming scheme  of Riemann  sheets.   }
\end{center}
\end{table}
In table~\ref{tableX} -- \ref{tableX4} we list several  fit results
with  different choices of ${\cal B}\sim \mathrm{a\,\,\,few}\times
10^{-4}$. The error of $\Gamma_c$ is given while others are not
listed.  For comparison we also list the fit results by setting
$\Gamma_c=0$. Notice that in tables~\ref{tableX} -- \ref{tableX4}
the parameter $g_X$ relates to parameter $g_1$ in
Eq.~(\ref{XBreit_wigner}) as $g_1={{g_X^2}\over
4\pi(m_{D^0}+m_{D^{*0}})^2}$.
\begin{table}[htp]
\begin{center}
\doublerulesep 0pt
\renewcommand\arraystretch{1.1}
\begin{tabular}{|c|c|c|c|c|c|}
\hline
 ${\cal B}=2\times 10^{-4}$ & $g_X$(GeV)  &  $E_f$(MeV)   &  $f_\rho\times 10^{3}$  &  $f_\omega\times10^{2}$&$\Gamma_c$(MeV) \\
    \hline
  $\chi^2_{eff}=4092$ &$4.16$  & $-6.79$&
 $2.10$ &   $1.45$& $1.78\pm1.66$ 
    \\
\hline

  $\chi^2_{eff}=4093$ &$4.40$  & $-6.40$&
 $0.44$ &   $0.32$& $-$

    \\

\hline
\end {tabular}
\caption {\label{tableX}  Pole locations:
$E_X^{III}=M-i\Gamma/2=-4.72- 1.51i$MeV,
$E_X^{II}=M-i\Gamma/2=-0.20- 0.38 i$MeV (with $\Gamma_c$);
$E_X^{III}=M-i\Gamma/2=-3.72- .08 i$MeV, 
$E_X^{IV}=M-i\Gamma/2=-0.02- 0.01 i$MeV (w/o $\Gamma_c$).} 
\end{center}
\end{table}
\begin{table}[htp]
\begin{center}
\doublerulesep 0pt
\renewcommand\arraystretch{1.1}
\begin{tabular}{|c|c|c|c|c|c|}
\hline
 ${\cal B}=3\times 10^{-4}$ & $g_X(GeV)$  &  $E_f$(MeV)   &  $f_\rho\times 10^{3}$  &  $f_\omega\times10^{2}$&$\Gamma_c$(MeV) \\
\hline

$\chi^2=4090$ &$4.20$  & $-6.89$&
 $1.46$ &   $1.01$& $2.02\pm 1.61$
    \\
\hline

$\chi^2=4092$ &$5.57 $  & $-10.3 $&
 $0.74$ &   $0.53$& $-$
    \\
\hline
\end {tabular}
\caption {\label{tableX1}  Pole positions:
$E_X^{III}=M-i\Gamma/2=-4.82-1.58 i$MeV,
$E_X^{II}=M-i\Gamma/2=-0.20- 0.40 i$MeV (with $\Gamma_c$);
$E_X^{III}=M-i\Gamma/2=-7.66- 0.12 i$MeV,
$E_X^{II}=M-i\Gamma/2=-0.02 -0.01 i$MeV (w/o $\Gamma_c$) } 
\end{center}
\end{table}
\begin{table}[htp]
\begin{center}
\doublerulesep 0pt
\renewcommand\arraystretch{1.1}

\begin{tabular}{|c|c|c|c|c|c|}

\hline
 ${\cal B}=5\times 10^{-4}$ & $g_X$(GeV)  &  $E_f$(MeV)   &  $f_\rho\times 10^{3}$  &  $f_\omega\times10^{2}$&$\Gamma_c$(MeV) \\
\hline

 $\chi^2=4088$&$5.41$  & $-11.1$&
 $2.05$ &   $1.39$& $3.23\pm2.54$
    \\
\hline $\chi^2=4091$ &$7.45$  & $-18.3 $&
 $1.67$ &   $1.18$& $-$
    \\
\hline
\end {tabular}
\caption {\label{tableX2} Pole positions: $E_X^{II}=-0.13- 0.39
i$MeV, $E_X^{III}=-9.20- 2.54 i$MeV (with $\Gamma_c$);
$E_X^{II}=+0.04- 0.08 i$MeV, $E_X^{III}=-18.9- 2.82 i$MeV (w/o
$\Gamma_c$).}
\end{center}
\end{table}
\begin{table}[htp]
\begin{center}
\doublerulesep 0pt
\renewcommand\arraystretch{1.1}
\begin{tabular}{|c|c|c|c|c|c|}
\hline
 ${\cal B}=1\times 10^{-3}$ & $g_X(GeV)$  &  $E_f$(MeV)   &  $f_\rho\times 10^{3}$  &  $f_\omega\times10^{2}$&$\Gamma_c$(MeV) \\
\hline

 $\chi^2=4086$  &$6.28$  & $-15.4$&
 $2.06$ &   $1.38$& $5.67\pm1.04$
    \\
    \hline

 $\chi^2=4090$  &$10.4$  & $-36.3 $&
 $3.49$ &   $2.44$& $-$
    \\
\hline
\end {tabular}
\caption {\label{tableX4}  $E_X^{II}=M-i\Gamma/2=-0.16-0.58 i$MeV,
$E_X^{III}=M-i\Gamma/2=-14.70- 4.30 i$MeV (with $\Gamma_c$);
$E_X^{II}=M-i\Gamma/2= -0.44-0.12 i$MeV,
$E_X^{III}=M-i\Gamma/2=-35.5- 1.40 i$MeV
 (w/o $\Gamma_c$)} 
\end{center}
\end{table}

By examining the  numerical results as  given in
tables~\ref{tableX}--\ref{tableX4} we  have the following
observations:
\begin{enumerate}

\item A third sheet pole is always found. When $\cal B$ gets large ($\sim 1\times 10^{-3}$), the pole locates far
away below the $D^0D^{*0}$ threshold. In such a case it is
understood that the whole data may well be fitted by a
parametrization with a single pole. In this sense, the $X(3872)$ may
be regarded as `dynamically generated'. However, for more reasonable
choices of
 (i.e.,
smaller) ${\cal B}$, the third sheet pole is rather close to the
threshold and is certainly physically relevant.

\item When $\cal B$ is small, the fit predicts a  value of
$\Gamma_c$ compatible with the quark model prediction on 2$^3P_1$
state light hadronic decay width, i.e., $\sim  1
$MeV~\cite{Meng07:X3872:decays}.

\item The location of the pole near $D^{*0}\bar D^0$ threshold is not stable in the
sense that it may either locate on sheet II or sheet IV. The former
corresponds to a $D^{*0}\bar D^0$ molecule, the latter corresponds
to a virtual state. The current analysis is not able to make a
definite conculsion on the two scenario, though the former is more
preferable.

\end{enumerate}
Besides above observations, in the fit when $\Gamma_c$ is set to
zero, we also confirm the approximate scaling law among different
parameters~\cite{hanhart}. However, when the constant width is
added, the approximate scaling law no longer exists. Because of the
approximate scaling law, the authors of Ref.~\cite{hanhart} fix one
of the parameters ($g$). Their choice
 is similar to the situation of table~\ref{tableX4}, corresponding to a large $\cal B$. Hence it
explains why in the analysis of Ref.~\cite{hanhart} the discussions
on the third sheet pole is missed, since the latter is quite distant
away from the physical region under concern. However, a choice of
${\cal B}\sim 1\times 10^{-3}$ seems to be too large to be
realistic.  A third sheet pole is of typical resonance behavior and
can be identified as the missing 2$^3P_1$ $c\bar c$ state.
  The  puzzle remained here is why the third sheet pole
locates below
 the $\bar D^0D^{*0}$ threshold. A pole
 with such a behavior is sometimes called  a ``crazy
 resonance".~\cite{taylor}

From tables 2--5 we notice that the location of the nearby pole is
not very stable numerically, though it seems to prefer to locate on
the second sheet (hence a molecule). The second sheet pole may
however shift above the $\bar D^0D^{*0}$ threshold, or even switch
to sheet IV. Therefore we hesitate to make any definite conclusion
on the location of this pole. The only solid statement that can be
drawn from above numerical analysis is that the third sheet pole
moves towards the $\bar D^0D^{*0}$ threshold and hence becomes
non-negligible when parameter $\cal B$ is within a few times
$10^{-4}$.

\begin{figure}[htbp]
\begin{flushleft}
{{\epsfig{file=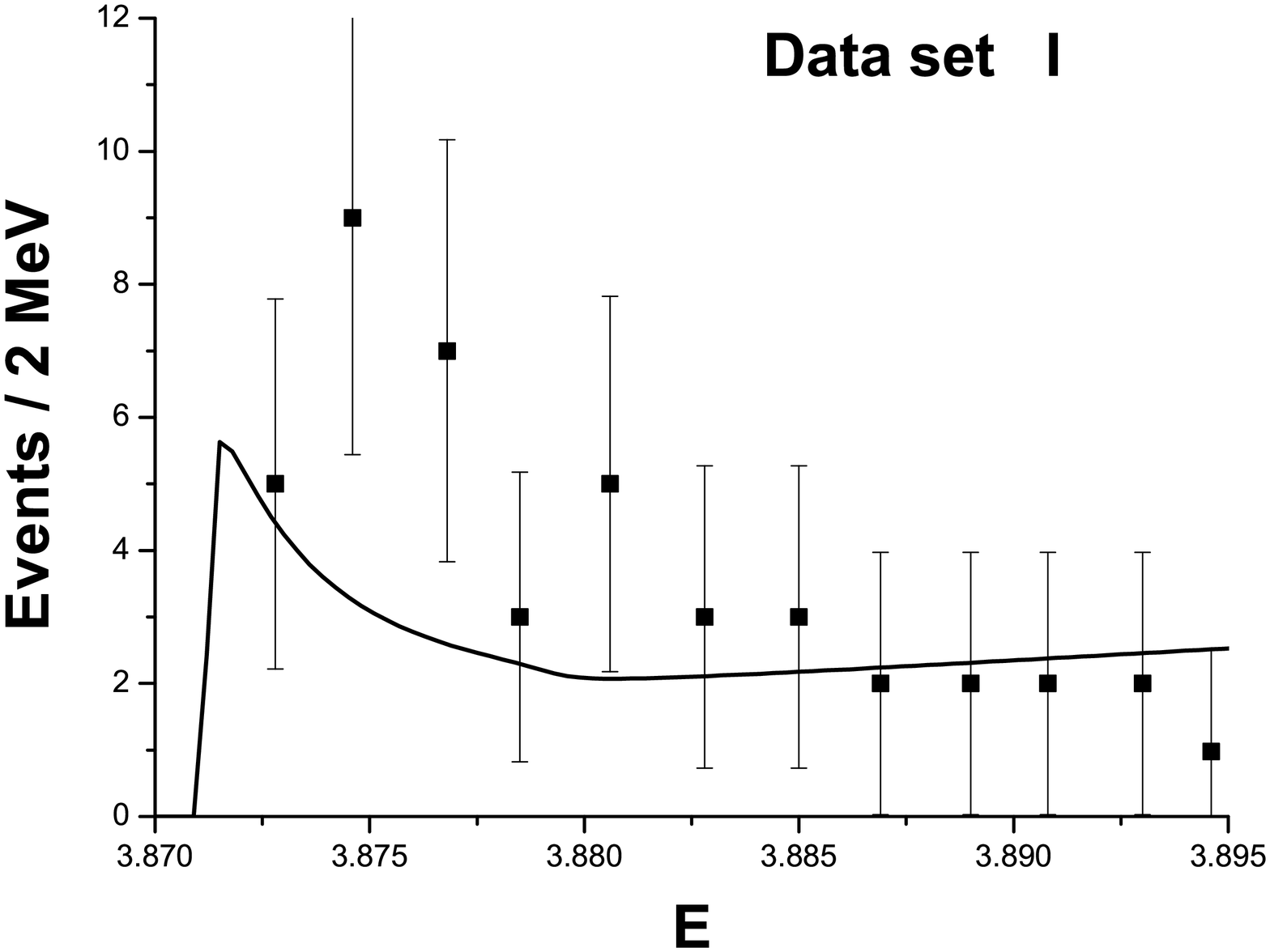,height=2.2in,width=2.8in}}}
\end{flushleft}
\vspace{-2.65in}
\begin{flushright}
{\mbox{\epsfig{file=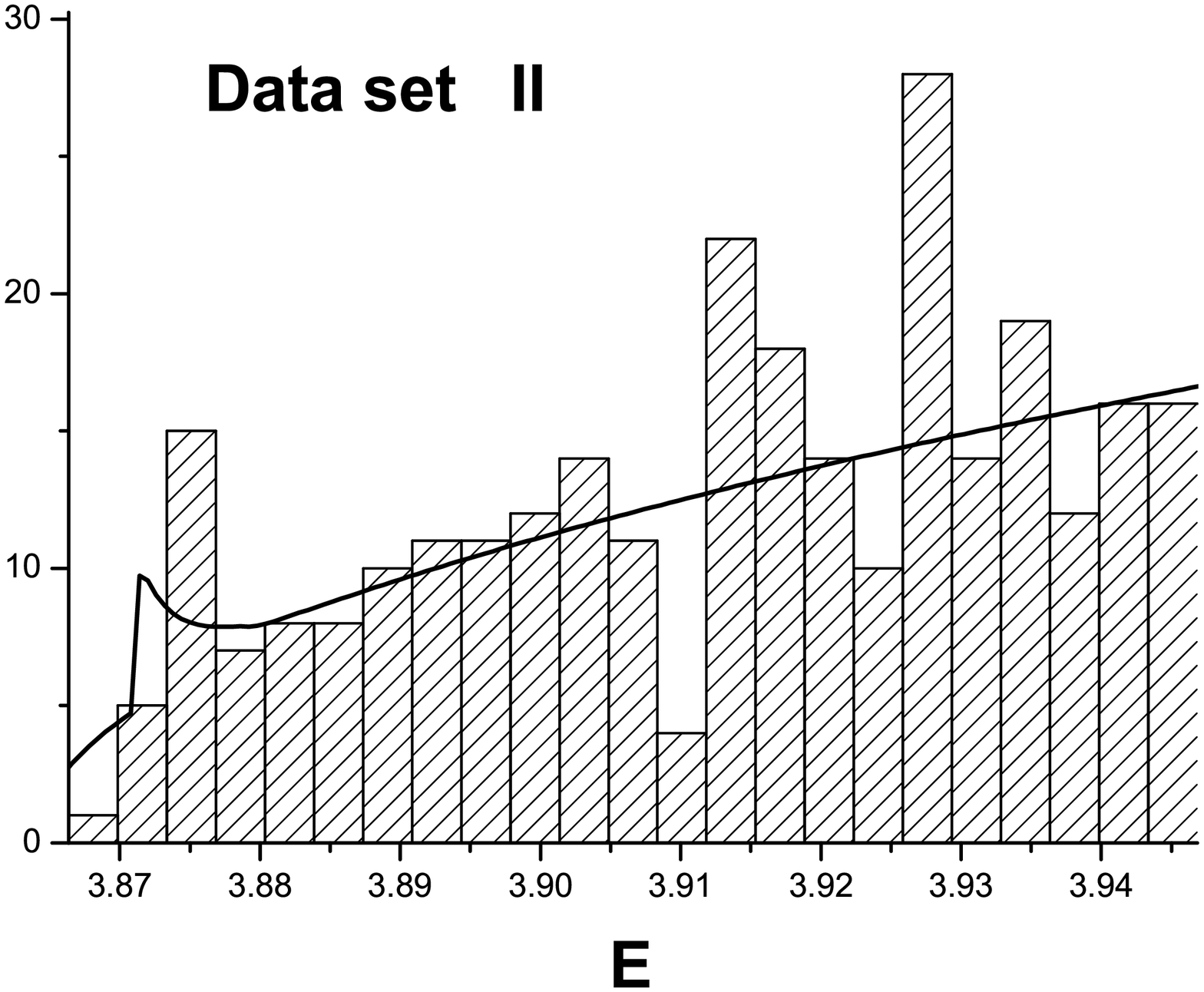,height=2.2in,width=2.8in}}}
\end{flushright}
\begin{flushleft}
{\mbox{\epsfig{file=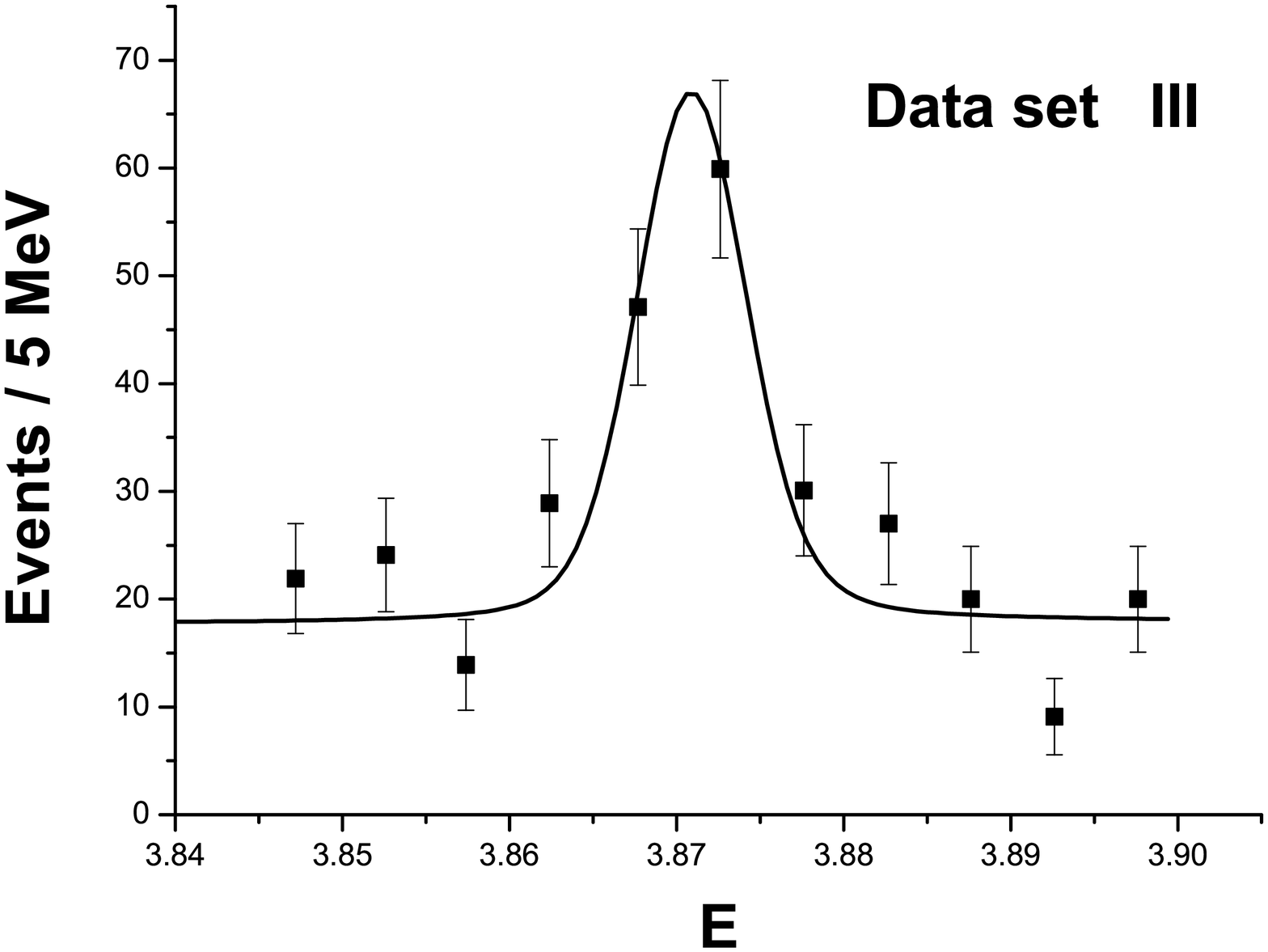,height=2.2in,width=2.8in}}}
\end{flushleft}
\vspace{-2.65in}
\begin{flushright}
{\mbox{\epsfig{file=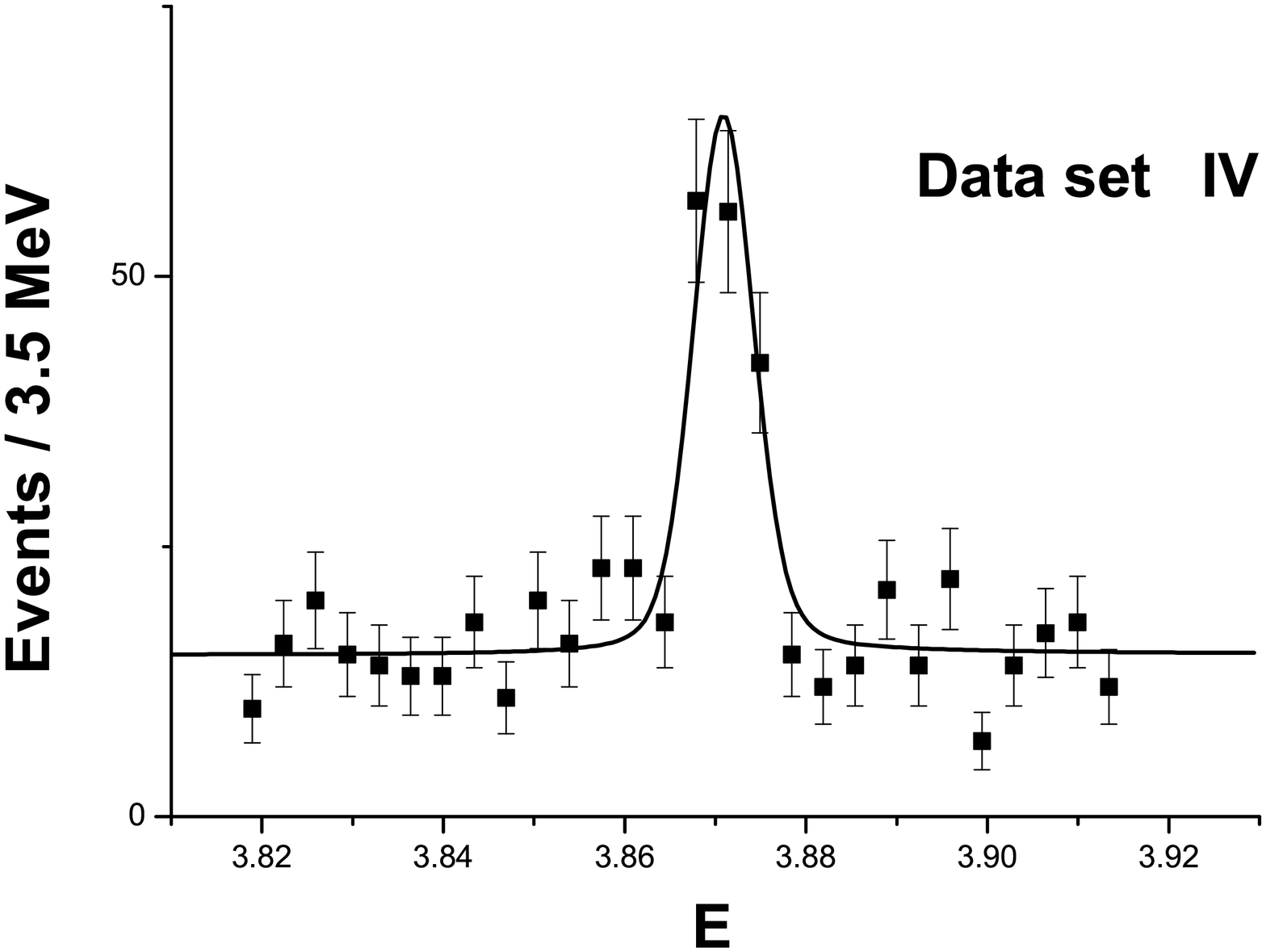,height=2.2in,width=2.8in}}}
\end{flushright}
\caption[]{Fit results with ${\cal B}=3\times 10^{-4}$, $\Gamma_c$
free. In order to compare with data samples II and IV, we set both
bin size to be 3.5MeV, and for the former we give the histogram.}
\label{fig3}
\end{figure}

In above  an effective minimization procedure with mixed $\chi^2$
function and likelyhood function is being used. To check weather the
qualitative picture revealed depends on the particular choice of
Eq.~(\ref{chisquareeffective}) or not, we also tested the binned
data fit by taking 1 bin=3.5MeV and fit to the same energy region.
It is found that the major conclusion of our qualitative result is
unchanged -- that a twin--pole structure is needed  when $\cal B$ is
small. Taking ${\cal B}=3\times 10^{-4}$ for example, the pole
locations are found to be: $E_X^{III}=M-i\Gamma/2=-3.84-1.71 i$MeV,
$E_X^{II}=M-i\Gamma/2=-0.10- 0.43 i$MeV (with $\Gamma_c$), to be
compared with the results of table~\ref{tableX1}.

The influence of different choices of the value of $R$ defined by
Eq.~(\ref{22}) is also tested. Setting $R=0.5$ and 1.5 for example,
it is found that, in all two cases, a  third sheet pole a few MeV
below $D\bar D^*$ threshold is  found, except the second sheet pole
very close  to the threshold. Therefore the variation of ratio $R$
does not distort the qualitative picture obtained in our numerical
analysis.

\section{Discussions and conclusions}
It is actually not surprising that our analysis finds two poles --
the occurrence of two poles is  an intrinsic character of the
coupled channel   Flatt\'e propagator. The importance of the current
analysis is that, as it points out, for reasonably chosen value of
 $\cal B$, the third sheet pole locates quite close to the
$D^0D^{*0}$ threshold and hence be physically relevant, except for
the sheet II (or sheet IV) pole emphasized in previous literature.
This picture is found to be unaltered when varying the fitting
method and the value of $R$. The conclusion certainly depends on the
simultaneous  fit to experimental data in two channels.  The
statistics of data set I and II are not as good as the
$J/\Psi\pi^+\pi^-$ data, hence future improvement on experimental
data in  $D^0\bar D^{0*}$ and $D^0\bar D^0\pi^0$ channels would
certainly be helpful in  clarifying the issue further.

 The two pole structure
of the $X(3872)$ state as revealed in this study is important, as we
believe, in understanding correctly the nature of the $X(3872)$
resonance. In this aspect, it can be  helpful to learn some lessons
from previous studies on the $f_0(980)$ resonance. Generally one
pole structure was considered as crucial evidence in supporting the
molecule identification of the $f_0(980)$ state in the literature.
On the other side, the existence of two poles close to the threshold
was often interpreted as an evidence against the molecular state
origin of the $f_0(980)$ resonance~\cite{Morgan92:pole-counting}.
  Early studies of the $f_0(980)$ tend to
identify it as having only one pole near the $\bar KK$ threshold,
and hence  a molecular state.~\cite{weinstein}
 It was found later  that the $\pi\pi,\bar KK$ scattering data are much
better described by allowing two poles near the $\bar KK$
threshold.\cite{penn,markushin}  In this picture, the third sheet
pole may contain a large $\bar qq$ component, that its position
close to the $\bar KK$ threshold is due to the attractive
interaction in the $\bar KK$ channel. The sheet II pole is mainly of
$\bar KK$ molecule nature. The $X(3872)$ situation should be rather
similar to the $f_0(980)$ case, except that in here the pole
locations are distorted more severely by coupled channel effects.
What we would like to stress here is that the two pole structure of
the state $X(3872)$ may reveal its dual faces: it is of $c\bar c$
origin due to the existence of the sheet III pole, but the coupled
channel effect also manifests itself by presenting an additional
pole, close to the $D^0D^{*0}$ threshold. The latter can also be
explained as molecular bound state/virtual state.

It should be stressed that, a pure molecular assignment of $X(3872)$
encounter a difficulty: the favored value of $\cal B$ lead to the
width $\Gamma_c$  to be roughly of $\mathcal{O}(1)$MeV. The pure
molecular
assignment of $X(3872)$, however, would predict a much smaller value
of $\Gamma_c$ as mentioned earlier.  Thus, our analysis supports
that $X(3872)$ is a mixing state of $\chi_{c1}'$ and
$D^0\bar{D}^{*0}$
components~\cite{Meng05:X3872:B-production,suzuki05:X3872:chi-c1-2P}.
A nearby $\chi_{c1}'$ below $D^{*0}\bar D^0$ threshold is actually
reported by quenched lattice QCD calculation.~\cite{chenying}
 The gap between the mass of
$\chi_{c1}'$ in the quark model~\cite{Barnes2005} and the
experimental one in Eq.~(\ref{M:X3872-CDF}) can be reduced when
coupled channel effect is taken into
account.~\cite{Li08:Couppled-vs-Screened} Here it is worth
emphasizing that the shift in the mass of a `pure' $\chi_{c1}'$ is
due to the attraction of the $D^0\bar{D}^{*0}$ threshold, not
because of its mixing with other $\bar cc$ state.

To conclude, the analysis given in this paper suggests the following
scenario for $X(3872)$: Firstly, there exists a sheet II (or sheet
IV) pole very close to the $D^{*0}\bar D^0$ threshold, this confirms
previous results in the literature. Secondly,  a fit to the data
with a reasonable choice of ${\cal B}$ parameter requires the
existence of a third sheet pole, but below  $D^{*0}\bar D^0$
threshold -- this observation is new. With the uncovering
 of the existence of two poles a clear understanding on the ambiversion of  $X(3872)$
 emerges --
 that it can be identified as a 2$^3P_1$ $\bar cc$ state strongly
 distorted by coupled channel effects.

\section{Acknowledgement}
We are grateful to our experimental colleagues, Yuan-Ning Gao and
Hai-Bo Li  for their kind helps and patient discussions. Especially
we are in debt to  Steve Olsen, who kindly provides us the original
Belle data, for helpful discussions and suggestions. It is also our
pleasure to thank Prof. K.~T.~Chao for helpful discussions. This
work is supported in part by National Nature Science Foundation of
China under Contract
nos. 10575002,
10721063, 
 and by China Postdoctoral Science Foundation under contract no.
20080430263.


\begin{thebibliography}{99}


\bibitem{Belle1}S.~K. Choi et al. (Belle Collaboration),
Phys. Rev. Lett. {\bf 91}(2003) 262001.

\bibitem{BaBar08:X3872:R_(0/+)}
B.~Aubert {\it et al.} [BaBar Collaboration], Phys.\ Rev.\ D {\bf
77}, 111101 (2008).

\bibitem{Belle08:X3872:R_(0/+)}
I.~Adachi {\it et al.} [Belle Collaboration], arXiv: 0809.1224
[hep-ex].

\bibitem{CDF08:X3872:MASS}
T.~Kuhr, talk given at QWG2008, Nara, Japan, see also the website:
http://www-cdf.fnal.gov/physics/new/bottom/080724.blessed-X-Mass.

\bibitem{Cleo07:D-mass}
C.~Cawlfield et al. [CLEO Collaboration], Phys. Rev. Lett. {\bf
98}(2007) 092002.


\bibitem{Belle05:X3872:psi-omega} K.~Abe et al. (Belle Collaboration),
hep-ex/0505037.

\bibitem{BaBar08:X3872:psi(')-gamma}
B.~Aubert {\it et al.} [BaBar Collaboration], arXiv: 0809.0042
[hep-ex].

%
%
%


\bibitem{Belle06:X3872:DDpi}G.~Gokhroo et al. (Belle Collaboration),
Phys. Rev. Lett. 97(2006)162002. 


\bibitem{BaBar08:X3872:DD*}B.~Aubert et al. (BaBar Collaboration), Phys. Rev. D77:
011102(2008). 



\bibitem{Belle08:X3872:DD*}
I.~Adachi {\it et al.} [Belle Collaboration], arXiv:0810.0358
[hep-ex].   We thank the Belle group for providing us with the data
points used in our fits.


\bibitem{Tornqvist:X3872:molecule}
 N.A. Tornqvist, Phys. Lett. B
{\bf 590}, 209 (2004); F. Close and P. Page, Phys. Lett. B {\bf
578}, 119 (2004); C.Y. Wong, Phys. Rev. C {\bf 69}, 055202 (2004);E.
Braaten and M. Kusunoki Phys.\ Rev.\ D {\bf 69}, 074005 (2004); M.B.
Voloshin, Phys. Lett. B {\bf 579}, 316 (2004).

\bibitem{Swanson04:X3872:molecule} E.S. Swanson, Phys. Lett. B {\bf 588}, 189 (2004); {\bf 598}, 197
( 2004).

\bibitem{Belle05CDF06:X3872:J^PC}
K.~Abe {\it et al.} [Belle Collaboration], arXiv: hep-ex/0505038; A.
Aulencia {\it et al.} [CDF Collaboration], Phy.\ Rev.\ Lett.\ {\bf
96}, 102002 (2006); {\bf 98}, 132002 (2007).


\bibitem{Meng05:X3872:B-production}
C. Meng, Y.J. Gao and K.T. Chao, arXiv: hep-ph/0506222.

\bibitem{suzuki05:X3872:chi-c1-2P}
M. Suzuki, Phys.\ Rev.\ D {\bf 72}, 114013 (2005).

\bibitem{Meng07:X3872:decays}
C. Meng and K.T. Chao, Phys.\ Rev.\ D {\bf 75}, 114002 (2007).

\bibitem{Morgan92:pole-counting}
D.~Morgan, Nucl.\ Phys.\ {\bf A543}, 632 (1992).

\bibitem{hanhart} C.~Hanhart, Yu.~S.~Kalashnikova, A.~E.~Kudryavtsev and
A.~V.~Nefediev, Phys. Rev. {\bf D76}, 034007 (2007).

\bibitem{braaten}E.~Braaten, M. Lu, Phys. Rev. {\bf D77}, 014029 (2008); Phys. Rev. {\bf D76}, 094028 (2007).

\bibitem{bugg}D.~Bugg, J.\ Phys.\ {\bf G35}, 075005 (2008).

\bibitem{oset}D.~Gammermann, E.~Oset, Eur. Phys. J. {\bf
A36}(2008)189; D.~Gamermann E.~Oset, Eur. Phys. J. {\bf
A33}(2007)119.

\bibitem{PDG08}
C.~Amsler et al. [Particle Data Group Collaboration], Phys.\ Lett.\
B {\bf 667}, 1 (2008).

\bibitem{Braaten:X3872:B-production}
E. Braaten, M. Kusunoki and S. Nussinov , Phys.\ Rev.\ Lett.\ {\bf
93}, 162001 (2004); E. Braaten and M. Kusunoki Phys.\ Rev.\ D {\bf
71}, 074005 (2005).

\bibitem{Molecule:XtoChicJ}
M.B.~Voloshin, Phys. Lett. B {\bf 604}, 69 (2004); S.~Dubynski and
M.B.~Voloshin, Phys.\ Rev.\ D {\bf 77}, 014013 (2008); S.~Fleming
and T.~Mehen, Phys.\ Rev.\ D {\bf 78}, 094019 (2008).

\bibitem{taylor}J. Taylor, \textit{Scattering Theory}, John Wiley
\& Sons, Inc., New York, 1972.

\bibitem{weinstein}J.~Weinstein and N. Isgur, Phys. Rev. Lett. {\bf
48}(1982)659; Phys. Rev. {\bf D27}(1983)588.

\bibitem{penn}K.~L.~Au, D.~Morgan and M.~R.~Pennington, Phys. Rev.
{\bf D35}(1987)1633; D.~Morgan and M.~R.~Pennington, Phys. Rev. {\bf
D48}(1993)1185.

\bibitem{markushin}M.~P.~Locher, V.~E.~Markushin, H.~Q.~Zheng, Euro.
Phys. J. {\bf C4}(1998)317.



\bibitem{chenying}Y.~Chen et al. (CLQCD Collaboration), arXive: hep-lat/0701021.

\bibitem{Barnes2005}
T.~Barnes, S.~Godfrey and E.S.~Swanson, Phys. Rev. D {\bf 72},
054026 (2005).

\bibitem{Li08:Couppled-vs-Screened}
B.Q.~Li, C. Meng and K.T. Chao, in preparation.

\end{thebibliography}
\end{document}